# A simulation-derived surrogate model for the vaporization rate of aluminum droplets heated by a passing shock wave

Pratik Das and H.S. Udaykumar

*Department of Mechanical Engineering, The University of Iowa, Iowa City, IA 52242*

## Abstract

The vaporization rate of aluminum droplets in shocked flows plays a crucial role in determining the energy release rate during the combustion of the aluminized energetic materials. In this paper, the physics of the vaporization of aluminum droplets in shocked flows is numerically investigated. Surrogate models for the temporally averaged Sherwood number and Nusselt number, cast as functions of shock Mach number and Reynolds number, are developed from the simulation-based data. The results show that the Sherwood number and the Nusselt number of the droplet increase monotonically with the Reynolds number. On the other hand, the Sherwood number and the Nusselt number exhibit non-monotonic behavior with increasing shock Mach number due to the transition of the post-shock flow from subsonic to the supersonic speeds as the shock Mach number is increased from 1.1 to 3.5. In contrast with available models in the literature that are commonly used in process scale computations of aluminum droplet vaporization, the current models for the Sherwood number and the Nusselt number are applicable over a wide range of the Reynolds number and the Mach number and will be useful in the macro-scale multi-phase simulations of the combustion of aluminumized energetic materials in high-speed flows.

## Nomenclature

$M_s$ = Incoming shock Mach number

$Re_D$ = Reynolds number

$Sh$ = Sherwood number

$Nu$ = Nusselt number

$\rho$ = Mixture density

$Y_i$ = Mass fraction of $i^{th}$ species

$\mathbf{u}$ = Velocity vector

$\mathbf{v}_i$ = Diffusion velocity of the $i^{th}$ species

$D_{i,mix}$ = Mixture-averaged diffusion coefficient of the $i^{th}$ species

$p$ = Pressure

$p_v$ = Partial pressure of aluminum vapor

$\mu$ = Viscosity

$E$ = Specific total energy

$T$ = Temperature

$k$ = Thermal conductivity

$\dot{q}''$ = Average heat-flux at the surface of the droplet

$\dot{\omega}''$ = Local evaporation mass-flux at the gas-liquid interface

$\dot{m}''$ = Average mass-flux per unit area at the droplet surface

$\overline{\dot{m}''}$ = Time-averaged vaporization mass flux at the droplet surface

$\overline{Nu}$ = Time-averaged Nusselt number

$\overline{Sh}$ = Time-averaged Sherwood number

$\phi$ = Levelset field

$\kappa$ = Local curvature at the gas-liquid interface

# 1 Introduction

The vaporization rate of the aluminum(Al) droplets plays a major role in determining the energy release rate during the combustion of the aluminized propellants and explosives. The after-burning of the Al droplets in solid rocket motors (SRMs)(Orlandi et al., 2019; Sabnis, 2003) and other high-speed multiphase flow applications such as particle-laden blast waves (Balakrishnan et al., 2012) enhances the rate of energy release in the system. The solid Al particles used in these applications undergo phase-change (melting followed by vaporization) prior to combustion. The vaporization of the Al droplets is a slower process than the combustion of the Al vapor in the gas phase. Therefore, the rate of energy release during the combustion of Al is limited by rate of vaporization of the Al droplets. The vaporization rate of the Al droplets is sensitive to the local flow-features such as the distribution of the pressure, temperature and velocity fields around the droplet, characterized by the Mach number($M_s$) and the Reynolds number($Re_D$). This paper studies the physics of aluminum droplet vaporization over a wide range of Reynolds numbers and Mach numbers, corresponding to a wide range of droplet size and flow speeds. Surrogate models are also constructed to quantify mass transport (Sherwood number) and heat transfer (Nusselt number) from droplets as functions of the parameters $Re_D$ and $M_s$. The present work significantly expands the parameter space over which the vaporization dynamics of Al droplets in high-speed flows has hitherto been understood and quantitatively modeled.

Physical experiments of vaporizing droplets in high-speed flows are sparse in the literature; comprehensive experimental studies have been typically focused on vaporizing droplets in low Mach number flows(Dai et al., 2019; Givler and Abraham, 1996; Lee and Law, 1992; Ranz and Marshal, 1952; Stengele et al., 1999). Among the few studies in high-speed compressible flows, Goosens et al.(Goossens et al., 1988) estimated the vaporization-timescale of droplets under shock-loading from experiments and found that the vaporization rate increases significantly with $M_s$. However, their study did not include the effects of droplet size and $Re_D$ on the vaporization rates. Most of the previous experimental studies on shock-droplet interaction are limited to characterizing the shock-induced breakup of droplets(Hirahara and Kawahashi, 1992; Hsiang and Faeth, 1995; Park et al., 2017; Sembian et al., 2016; Yoshida and Takayama, 1990) and measuring the attenuation of a shock-wave passing through a cluster of droplets(Borisov et al., 1971; Mataradze et al., 2019). Accurate calculation of the vaporization rate of the droplets in shocked flows from experiments still remains challenging and expensive because of the small time scales and the length scales of flow and thermal transport processes.

## 1.1 Computational models for droplet vaporization in high-speed flows

As an alternative to experimental approaches, computational studies offer the ability to visualize and quantify the vaporization mechanisms of micron-sized droplets in high-speed flows. Over the past three decades, several numerical methods have been developed for the interface-resolved direct numerical simulations of multiphase flows. Among them, sharp-interface methods such as front-tracking(Tryggvason et al., 2001), Volume of Fluid(Hirt and Nichols, 1981) and levelsets (Sethian and Smereka, 2003), have been shown to be promising in accurately computing the vaporization rate of droplets. A major challenge for these numerical methods is to couple the flow-fields of the gas and the liquid phases at the interface.

The numerical modelling of the interfacial mechanics and thermodynamics in the compressible flow regime must capture the interaction of the shocks and expansion waves with the interface. The flow-fields in the two phases at the interface must be coupled such that the characteristic waves can propagate from one material to the other across the interface while respecting the interfacial jump conditions due to the surface tension and phase-change. To this end, the Riemann-solver based ghost fluid method(rGFM)(Liu et al., 2003; Sambasivan and Udaykumar, 2009) has been successfully used in the sharp-interface calculations of inviscid multiphase flows in the compressible regime. In rGFM, a 1-D local Riemann problem is solved at the interface to couple the flowfield in the materials separated by the interface. For interfaces undergoing phase change, the rGFM must satisfy the interfacial jump-conditions due to surface tension, vaporization and the jump in the viscous stresses. Such robust rGFM methods capable of handling shock interactions with vaporizing gas-liquid interface have been developed only recently (Das and Udaykumar, 2020; Houim and Kuo, 2013). The interface resolved simulations have not yet been used to study the effects of the incoming shock strength($M_s$) on the vaporization rate of the droplets in shocked flows.

Such interface-resolved calculation of shock-droplet interactions can be computationally expensive. Numerical calculations involving only a few droplets(~O(10)-~O(100)) can now be performed on a modern supercomputer. However, process-scale simulations of the internal flow in an SRM involve millions of Al droplets; resolving the dynamics of shock-droplet interaction in such a calculations is impractical. A computationally tractable approach for solving such process-scale problems is to model the interaction of the continuous gaseous phase and the dispersed liquid phase. In such multiphase models, the continuous gaseous phase is modeled in an Eulerian frame of reference, while the liquid droplets can be modeled either as one phase in a continuous multi-phase mixture in the Eulerian frame of reference (Enwald et al., 1996; Houim and Oran, 2016; Shotorban et al., 2013) or as dimensionless points in the Lagrangian frame of reference (Jacobs et al., 2012; Jacobs and Don, 2009). Such modeling strategies for multiphase systems are classified as the Eulerian-Eulerian (Balakrishnan et al., 2012; Houim and Oran, 2016; Saito, 2002) methods and the Eulerian-Lagrangian (Dahal and McFarland, 2017; Davis et al., 2017; Jacobs and Don, 2009; Sabnis, 2003) methods respectively. The Eulerian-Eulerian or Eulerian-Lagrangian models rely on closure terms in the mass, momentum, and energy conservation equations to model the interaction of the continuous and dispersed phases. For example, the closure term in the mass conservation equation in these models account for the mass exchange between the gaseous and the dispersed liquid phase due to vaporization. The mass exchange between the gas and the liquid droplets is modeled as a function of the Sherwood number($Sh$)(Balakrishnan et al., 2012; Dahal and McFarland, 2017; Sabnis, 2003). Similarly, the momentum and heat transfer between the gas and the droplets are modeled in terms of the Nusselt number ($Nu$) and the drag coefficient($C_D$), respectively. The accuracy of the macro-scale multi-phase flow models depends on the accuracy of the closure models for interphase mass, momentum and energy transfer.

## 1.2 Closure models for the thermo-mechanics of gas-liquid mixtures in high-speed flows

Typically, closure models are provided in terms of empirical correlations of $Sh$ and $Nu$ as functions of $Re_D$. For example, Dahal and McFarland(Dahal and McFarland, 2017) used the Ranz-Marshall model for $Sh$ in their macroscale calculation of shock-interaction with vaporizing droplets. Another empirical model(Kreith et al., 2012) for the $Sh$ and $Nu$ correlations was used in (Sabnis, 2003) for the calculation of aluminized propellant combustion in SRMs. Smirnov et al. (Betelin et al., 2012; Smirnov et al., 2013) used an empirical correlation for $Nu$ to model the heat and mass transfer between the evaporating droplets and the gas-phase. Such empirical or heuristic models for $Sh$ and $Nu$ are obtained under assumptions of incompressible flows and are developed through experiments in a limited range of the relevant parameter space. In fact, current models for $Sh$ available in the literature(Abramson and Sirignano, 1989; Kreith et al., 2012; Ranz and Marshal, 1952) are adapted from data in the incompressible flow regime. These models for $Sh$ do not explicitly account for the effects of the shock-compression of the gas around the droplet on the vaporization rate of droplets interacting with shocks.

The current paper is the first attempt to study the effects of $M_s$ on the $Sh$ of Al droplets using interface-resolved numerical simulations and develop correlations for $Sh$ and $Nu$ as functions of $M_s$ and $Re_D$ from the simulation derived data. In the current simulations of shock-droplet interaction, the gas-liquid interface of the droplet is explicitly tracked using the levelset method(Das and Udaykumar, 2020; Sethian and Smereka, 2003) and interfacial mechanics is captured using the rGFM developed in(Das and Udaykumar, 2020). The vaporization rate at the droplet interface is calculated from the Schrage-Knudsen vaporization model(Schrage, 1953) derived from kinetic theory of gases directly. Therefore, in the current simulations, the effects of the variation of pressure along the interface of droplets is explicitly accounted for while computing the vaporization rate of the droplets. As a result, the $Sh$ and $Nu$ of the Al droplets computed in the current simulations takes the effects of non-linear wave interactions at the interface on the mass and heat transfer rates into account.

A single droplet is considered for tractable resolved simulations of shock-droplet interaction to calculate the *Sh* and the *Nu*. The simulation data is then used to develop surrogate models for *Sh* and *Nu* as functions of $M_s$ and $Re_D$. The surrogate models are developed using a modified Bayesian Kriging(MBKG) based multiscale-surrogate modeling technique as demonstrated in the previous works(Sen et al., 2015, 2017a). Previously, the MBKG method was used to develop surrogate models for $C_D$(Das et al., 2018a; Sen et al., 2015, 2017a) and $Nu$ (Das et al., 2018b)from interface resolved calculations of shock-particle interactions where the particles remained solid, i.e. adiabatic conditions were assumed at the solid-gas interface. In this work, the MBKG method (Das et al., 2018b; Gaul, 2014a; Sen et al., 2015, 2017a) is used to develop the surrogate models for *Sh* and *Nu* for shock interaction with droplets.

The remainder of the paper is organized as follows. The sharp-interface method used in the current calculations of shock interaction with the vaporizing droplets is discussed in section 2. The current method is verified against the exact solution for an air-liquid Al shock-tube problem in section 3.1. The different stages of shock interaction with a cylindrical Al droplet are discussed in Section 3.2. The effect of $M_S$ and $Re_D$ on the *Sh* and the *Nu* of an Al droplet during the interaction with a shock wave is studied in Section 3.3. Surrogate models for *Sh* and Nu of the droplet are developed from the simulation-based data in Section 3.4. The concluding remarks from the current work are in section 4.

## 2  Methods

A Cartesian grid-based sharp-interface method is used to compute the vaporization of Al droplets in shocked flows. The sharp interface between the gaseous and the liquid phases at the surface of the droplet is tracked using the levelset method. The embedded interface divides the flowfield into the gaseous and the liquid phase. The governing equations for compressible multicomponent flows in Cartesian co-ordinates, outlined below, are solved in each phase to evolve the flowfield in time.

### 2.1  Governing Equations:

The mass conservation equation for the $i^{th}$ species in each phase is given by:

$$\frac{\partial}{\partial t}(\rho Y_i) + \nabla \cdot (\rho \boldsymbol{u} Y_i) = \nabla \cdot (\rho Y_i \boldsymbol{v}_i) + \dot{\omega}_i \tag{1}$$

where $\rho$ and $\boldsymbol{u}$ are the mixture density and velocity field, respectively. $Yi$ and $\boldsymbol{v}_i$ are the mass-fraction and the diffusion velocity of the $i^{th}$ species respectively. Here the gaseous phase has two components, air ($i = 1$) and Al vapor ($i = 2$). The liquid phase is treated as a pure Al ($i = 1$ in the liquid phase).

The diffusional velocity $\boldsymbol{v}_i$ is calculated as follows to ensure the conservation of mass(Coffee and Heimerl, 1981):

$$\boldsymbol{v}_i = \hat{\boldsymbol{v}}_i - \sum_j Y_j \hat{\boldsymbol{v}}_j \tag{2}$$

where $\hat{\boldsymbol{v}}_i$ is calculated from:

$$\hat{\boldsymbol{v}}_i = -\frac{D_{i,mix}}{X_i} \nabla X_i \tag{3}$$

$X_i$ and $D_{i,mix}$ are the mole-fraction and the mixture-averaged diffusion coefficient of the $i^{th}$ species. $D_{i,mix}$ is calculated from binary diffusion co-efficient $D_{ij}$ using the following equation(Kee et al., 2003):

$$D_{i,mix} = \frac{1 - Y_i}{\sum_{j, j \neq i} X_i / D_{ij}} \tag{4}$$

$D_{ij}$ of air and Al vapor is calculated using the Chapman-Enskog theory(Kee et al., 2003).

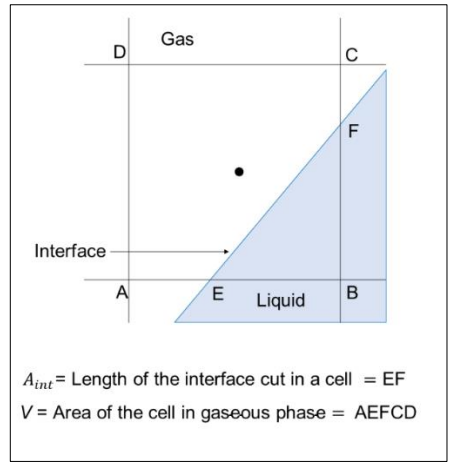

Figure 1 Calculation of the rate of vaporization at the gas-liquid interface

The source-term $\dot{\omega}_i$ in Eq. (1) is the rate of production of $i^{th}$ species in the gaseous phase at the interface. Therefore, $\dot{\omega}_i$ is non-zero only at the computational cells adjacent to the interface. $\dot{\omega}_i$ is computed as follows:

$$\dot{\omega}_i = \begin{cases} 0, for\ i = 1 \\ \dfrac{\dot{\omega}" \times A_{int}}{V}, for\ i = 2 \end{cases} \tag{5}$$

where $\dot{\omega}"$ is the local evaporation mass-flux at the droplet surface. $A_{int}$ is the length(area in 3D) of the interface within a computational cell and $V$ is the area(volume in 3D) occupied by the gaseous phase in the computational cell. The notion of $A_{int}$ and $V$ are demonstrated in Figure 1 for a 2D scenario. $A_{int}$ and $V$ are computed using algorithms described in previous work (Mousel, 2012; Scardovelli and Zaleski, 2000).

A non-equilibrium vaporization models is used in this work to calculate $\dot{\omega}"$ at the droplet surface. The previous studies on the comparison of the equilibrium and non-equilibrium vaporization models in (Miller et al., 1998; Tyurenkova, 2012) show that the thermodynamically non-equilibrium effects become important for the smaller droplets($D \sim 1 \mu m$ ) and the results obtained using the equilibrium and non-equilibrium assumptions deviate from each other with the decrease in the droplet size. In this work, a transient problem of droplet vaporization under the influence of shock is solved and droplet sizes considered in the current work fall within the range of $0.1 \mu m - 20 \mu m$ in diameter. Under such conditions, the assumption of phase equilibrium at the droplet interface does not hold. The following Schrage-Knudsen non-equilibrium vaporization model (Houim and Kuo, 2013; Schrage, 1953) is used in this work to calculate $\dot{\omega}"$:

$$\dot{\omega}" = \dfrac{2C}{2-C} \sqrt{\dfrac{Mw_{Al}}{2\pi R_u}} \left( \dfrac{p_{sat}}{\sqrt{T_l}} - \dfrac{p_v}{\sqrt{T_g}} \right) \tag{6}$$

where

$$C = \left\{ 1 - \left(\dfrac{\rho_g}{\rho_l}\right)^{\frac{1}{3}} \right\} exp\left( -\dfrac{1}{2(\rho_l/\rho_g)^{1/3} - 2} \right)$$

$T_l$ and $T_g$ are the temperature of the liquid and the gaseous phase at the interface. $p_{sat}$ is the saturation vapor pressure of Al at $T_l$ and $p_v$ is the local partial vapor pressure of Al at the interface. $R_u$ and $Mw_{Al}$ are the universal gas constant and the molecular weight of Al respectively.

To compute $p_{sat}$ for the curved surfaces, first the saturation vapor pressure of aluminum at a planner interface $(p_{sat,planner})$ of the gaseous and liquid aluminum is computed from the following equation(Gathers, 1983; Houim, 2011):

$$p_{sat,planner}(Pa) = exp\left(36.547 - \frac{39033}{T_l} - 1.3981\, ln(T_l) + 6.7839 \times 10^{-9} T_l^2\right) \tag{7}$$

The $p_{sat}$ for a curved surface is computed from the Kelvin equation to incorporate the effect of curvature of the surface and the surface-tension as follows:

$$p_{sat} = p_{sat,planner}\, exp\left(\frac{2\gamma Mw_{Al}}{r\rho_l R_u T}\right) \tag{8}$$

where $r$ is the local radius of curvature at a given location on the interface. $\gamma$ is the surface-tension of liquid aluminum. $\gamma$ is expressed as a function of $T_l$ as follows(Houim, 2011; Sarou-Kanian et al., 2003):

$$\gamma(N/m) = 1.2796 - 0.000274 T_l(K) \tag{9}$$

$p_v$ is calculated as follows:

$$p_v = \frac{Y_{al}\rho R_u T_g}{Mw_{Al}} \tag{10}$$

The momentum conservation equation is given by:

$$\frac{\partial}{\partial t}(\rho \mathbf{u}) + \nabla \cdot (\rho \mathbf{u}\mathbf{u}) = -\nabla p + \nabla \cdot \left\{\mu(\nabla \mathbf{u} + \nabla \mathbf{u}^T) - \frac{2}{3}\mu \nabla \cdot \mathbf{u}\right\} + \mathbf{M} \tag{11}$$

where $p$ is the pressure. In the gaseous phase, $\mu$ is the mixture viscosity and calculated as given in(Houim, 2011; Kee et al., 2003). The viscosity of air is calculated from Sutherland's law given by:

$$\mu_{air} = \frac{\mu_{ref}\left(\frac{T}{T_{ref}}\right)^{\frac{3}{2}}(T_{ref}+S)}{T+S} \quad (12)$$

where $\mu_{ref} = 1.813 \times 10^{-5}$ Pa.s, $T_{ref} = 293.0$K and $S = 110.4$

$\mu$ of Al in liquid phase is calculated in the S.I. units from(Assael et al., 2006):

$$\mu = 1.85183 \times 10^{-4} e^{\frac{1850.1}{T}} \quad (13)$$

where $T$ is the temperature.

The source terms $\boldsymbol{M}$ represents the momentum-exchange between the gas and the liquid phase due to the phase-change at the interface. $\boldsymbol{M}$ is calculated from:

$$\boldsymbol{M} = \sum_i \dot{\omega}_i \boldsymbol{u}_{I,g} \quad (14)$$

where $\boldsymbol{u}_{I,g}$ is the velocity of the gaseous phase at the interface.

The conservation-law for the total specific-energy $E$ is given by:

$$\frac{\partial}{\partial t}(\rho E) + \nabla \cdot \{\boldsymbol{u}(\rho E + p)\} = \nabla \cdot \left\{\nabla(kT) - \sum_i h_i \rho Y_i \boldsymbol{v}_i\right\} + S_V + S_E \quad (15)$$

where $E$ is the specific total energy. In the gaseous phase, $k$ is the mixture averaged thermal-conductivity and computed as given in (Houim, 2011; Kee et al., 2003). $k$ in air is calculated from:

$$Pr = \frac{C_p \mu}{k} = 0.72 \quad (16)$$

where $Pr$ is the Prandlt number and $C_p$ is the specific heat capacity of air at constant pressure. $k$ for Al in the liquid-phase is calculated from the following equation in the S.I. units(Recoules and Crocombette, 2005):

$$k = 42 + 0.056T \tag{17}$$

$h_i$ is the specific enthalpy of the $i^{th}$ species at the temperature $T$, calculated as follows:

$$h_i = h_{f,i} + \int_{T^o}^{T} C_{p,i}(\tau)d\tau \tag{18}$$

where $h_{f,i}$ is the specific enthalpy of formation of the $i^{th}$ species at the reference state ($T^0$=298K). $C_{p,i}(T)$ is the specific heat capacity at constant pressure of the $i^{th}$ species, at a temperature $T$. $C_{p,i}(T)$ is a polynomial function of temperature $T$ in the absolute scale and taken from (Burcat, 1984).

$S_V$ in Eq. (15) accounts for the energy source due to the viscous dissipation, given by:

$$S_v = \nabla \cdot \left\{ \boldsymbol{u} \cdot \left\{ \mu(\nabla \boldsymbol{u} + \nabla \boldsymbol{u}^T) - \frac{2}{3}\mu \nabla \cdot \boldsymbol{u} \right\} \right\} \tag{19}$$

The source terms $S_E$ represent the energy exchange between the gas and the liquid phase at the interface. $S_E$ is calculated from:

$$S_E = \sum_i \dot{\omega}_i \left\{ h_i - \frac{R_u}{Mw_i}T \right\} \tag{20}$$

The pressure $p$ is computed from the equations of state. Separate equations of state are used in the gaseous and liquid phases. $p$ in the gaseous phase is calculated by applying Dalton's law for ideal gases:

$$p = \sum_i p_i = \rho R_u T \sum_i \frac{Y_i}{Mw_i} \tag{21}$$

where $p_i$ is the partial pressure of the $i^{th}$ component of the gaseous mixture. In the gaseous phase, $T$ is obtained by solving:

$$E(T) = \sum_i \left( Y_i h_i - \frac{R_u}{Mw_i}T \right) + \frac{u^2 + v^2 + w^2}{2} \tag{22}$$

The Tait EOS in the following form is used to obtain $p$ in the liquid phase:

$$p = B\left\{\left(\frac{\rho}{\rho_0}\right)^N - 1\right\} + A \quad (23)$$

where, $A$, $B$, $N$ and $\rho_0$ are physical constants shown in the Table 1.

| $A(Pa)$ | $B(Pa)$ | $N$ | $\rho_0(kg/m^3)$ |
|---|---|---|---|
| $10^5$ | $3.36 \times 10^9$ | 8.55 | 2003.0 |

Table 1 Values of the physical parameters in the Tait EOS for liquid Al

$E$ is related to $T$ through the following equation (Houim and Kuo, 2013):

$$E = E_{ref} + C_v(T - T_{ref}) + \frac{u^2 + v^2 + w^2}{2} \quad (24)$$

where $E_{ref}$ and $T_{ref}$ are the reference specific energy and temperature of the liquid Al. $E_{ref}$ is calibrated at $T_{ref} = 2743.0$ K for liquid Al in this work.

The governing equations of the multi-component compressible flows described above are solved independently in the gaseous and the liquid phases which are separated by the embedded sharp interface within the computational domain.

### 2.2 Interface tracking:

The sharp interface between the gaseous and the liquid phase is tracked using the levelset method(Osher and Sethian, 1988; Sethian and Smereka, 2003). The zero levelset contour defines the location of the sharp interface between the liquid and the gaseous phases. A narrow-band levelset field provides the signed normal distance to the nominal interface from any point in a band around the sharp interface. The levelset field is advected to capture the evolution of the interface as the flow evolves in time:

$$\frac{\partial \phi}{\partial t} + \boldsymbol{u_n} \cdot \nabla \phi = 0 \quad (25)$$

where $\phi$ represents the levelset field. $\boldsymbol{u_n}$ is the local normal velocity of the interface. $\boldsymbol{u_n}$ is computed from the following equation:

$$\boldsymbol{u_n} = -\frac{\dot{\omega}''}{\rho_l}\boldsymbol{n} + \boldsymbol{u_l} \qquad (26)$$

where $\boldsymbol{u_l}$ is the velocity of the liquid phase at the interface, $\boldsymbol{n}$ is the local unit vector normal to the interface and $\rho_l$ is the local density of the liquid.

## 2.3 Interfacial Treatment:

The gas and the liquid flow-fields are coupled at the sharp interface using the ghost fluid method(GFM)(Fedkiw et al., 1999) such that the following interfacial jump conditions are satisfied:

$$[u_n] = \dot{\omega}'' \left[\frac{1}{\rho}\right] \qquad (27)$$

$$[p] = -\gamma\kappa - \dot{\omega}''[u_n] - \left[2\mu\frac{\partial u_n}{\partial n} - \frac{2}{3}\mu\left(\frac{\partial u_n}{\partial n} + \frac{\partial u_s}{\partial s}\right)\right] \qquad (28)$$

$$\left[\mu\left(\frac{\partial u_s}{\partial n} + \frac{\partial u_n}{\partial s}\right)\right] = -\frac{d\gamma}{ds} \qquad (29)$$

$$[\dot{q}''_{cond}] = -\dot{\omega}''[h] + \left[\left\{2\mu\frac{\partial u_n}{\partial n} - \frac{2}{3}\mu\left(\frac{\partial u_n}{\partial n} + \frac{\partial u_s}{\partial s}\right)\right\}u_n\right] + \left[\left\{\mu\left(\frac{\partial u_s}{\partial n} + \frac{\partial u_n}{\partial s}\right)\right\}u_s\right] \qquad (30)$$

where the operator [ ] represents the jump, as in:

$$[\chi] = \chi_g - \chi_l$$

$\chi$ is any flow variable of interest. The subscripts $g$ and $l$ represent the flow variables at the interface in the gaseous and the liquid phase respectively. The subscripts $n$ and $s$ represent the directions normal and tangential to the interface. $\gamma$ is the local surface tension at the gas-liquid interface. $\kappa$ is the local curvature at the interface and is calculated from the levelset field(Sethian and Smereka, 2003) using the following equation:

$$\kappa = \nabla \cdot \left(\frac{\nabla\phi}{|\nabla\phi|}\right) \qquad (31)$$

Eq. (27) accounts for the jump in the normal velocity of the two phases at the interface caused by vaporization. The pressure jump in Eq. (28) at the interface is due to surface-tension $(-\gamma\kappa)$, vaporization $(\dot{\omega}"[u_n])$ and jump in the normal component of viscous stress $\left(\left[2\mu\frac{\partial u_n}{\partial n} - \frac{2}{3}\mu\left(\frac{\partial u_n}{\partial n} + \frac{\partial u_s}{\partial s}\right)\right]\right)$. The jump in the tangential components of the deviatoric stress tensor $\left(\left[\mu\left(\frac{\partial u_s}{\partial n} + \frac{\partial u_n}{\partial s}\right)\right]\right)$ in the Eq. (29) represents the effect of Marangoni stresses at the interface. The jump in the heat flux $\left(\left[\dot{q}"_{cond}\right]\right)$ is given by Eq. (30). It accounts for the latent heat of evaporation $(\dot{\omega}"[h])$ and the work done by the viscous stresses.

The interfacial jump-conditions in Eq.(27)-(30) are integrated with a Riemann-solver based GFM(RS-GFM) to ensure accurate coupling of the characteristic waves in the gaseous and the liquid phases at the interface. Details of the modified RS-GFM can be found in previous work(Das and Udaykumar, 2020).

## 2.4 Discretization Schemes:

An operator-splitting algorithm is used to perform time integration of the governing equations, Eq. (1) -(15). The governing equations for species, momentum, and the energy-conservation are split into the hyperbolic, parabolic and the source terms as specified in the following table:

|  | Hyperbolic terms | Parabolic terms | Source terms |
|---|---|---|---|
| Species conservation eqn. | $\nabla \cdot (\rho \boldsymbol{u} Y_i)$ | $\nabla \cdot (\rho Y_i \boldsymbol{v}_i)$ | $\dot{\omega}_i$ |
| Mom. conservation eqn. | $\nabla \cdot (\rho \boldsymbol{u}\boldsymbol{u}) - \nabla p$ | $\nabla \cdot \left\{\mu(\nabla \boldsymbol{u} + \nabla \boldsymbol{u}^T) - \frac{2}{3}\mu \nabla \cdot \boldsymbol{u}\right\}$ | $\boldsymbol{M}$ |
| Energy conservation eqn. | $\nabla \cdot \{\boldsymbol{u}(\rho E + p)\}$ | $\nabla \cdot \left\{\nabla(kT) - \sum_i h_i \rho Y_i \boldsymbol{v}_i\right\} + S_V$ | $S_E$ |

Table 2 The hyperbolic, parabolic and the source terms in the governing equations Eq. (1) -(15)

The hyperbolic terms in the governing equations are first integrated using a third-order Runge-Kutta (TVD-RK) scheme(Gottlieb and Shu, 1998) to obtain an intermediate solution state $\boldsymbol{U}^*(= [\rho^*, \boldsymbol{u}^*, E^*])$ at the n$^{\text{th}}$ timestep:

$$\boldsymbol{U}^* = H^{\Delta t}(\boldsymbol{U}^n) \tag{32}$$

where $U^n$ is the solution state at the end of the n$^{th}$ time-step. $H^{\Delta t}()$ is the linearized hyperbolic operator. The parabolic terms in the governing equations are integrated using the Runge-Kutta-Chebyshev(RKC) explicit time integration scheme(Verwer et al., 2004) to obtain a second intermediate state $U^{**}$ from $U^*$ :

$$U^{**} = P^{\Delta t}(U^*) \qquad (33)$$

where $P^{\Delta t}()$ is the parabolic operator. Finally, the source terms are integrated using 4$^{th}$ order Runge-Kutta-Fehlberg scheme to obtain the solution at the n+1$^{th}$ time step:

$$U^{n+1} = S^{\Delta t}(U^{**}) \qquad (34)$$

The time-step size $\Delta t$ is dependent on the hyperbolic operator and obtained from the CFL number:

$$\Delta t = CFL \left[\frac{\Delta x}{u+a}\right]_{min}, \text{ where CFL} \leq 1, \Delta x \text{ is grid size and a is the wave speed} \qquad (35)$$

A 3$^{rd}$ order accurate ENO-LLF(Shu and Osher, 1989) scheme is used for spatial discretization of the hyperbolic terms in the governing equations. A 4$^{th}$-order accurate finite difference scheme(Das et al., 2017) is used to discretize the parabolic terms. Further details of the numerical schemes can be found in previous work(Das et al., 2017; Das and Udaykumar, 2020).

## 2.5 Calculation of the quantities of interest:

The sharp-interface method described above is used to perform simulations of shock interactions with vaporizing droplets over a range of shock Mach numbers($M_s$) and Reynolds numbers($Re_D$). In these simulations, the post-shock conditions are selected as the reference to define the Nusselt number($Nu$) and Sherwood number($Sh$).

$Re_D$ is defined as:

$$Re_D = \frac{\rho_{ps} u_{ps} D}{\mu} \qquad (36)$$

where $\rho_{ps}$ and $u_{ps}$ are the post-shock density and velocity of air. $D$ is the diameter of the droplet and $\mu$ is the viscosity of air.

*Nusselt number for a single droplet*

Similarly, the instantaneous Nusselt number($Nu$) for a single droplet is calculated from:

$$Nu(t) = \frac{\dot{q}"(t)D}{k_{air}(T_p - T_{ps})} \tag{37}$$

where $T_p$ is the average temperature of the droplet and $T_{ps}$ is the temperature of air behind the shock. $\dot{q}"(t)$ is the average heat-flux at the surface of the cylindrical droplet and it is calculated as:

$$\dot{q}"(t) = \frac{\oint_S \boldsymbol{q_n} \cdot \boldsymbol{n} dl}{S} \tag{38}$$

where $\boldsymbol{q_n}$ is the local heat-flux at any point on the interface. $dl$ is the infinitesimal segment of the liquid droplet surface and $S$ is the total length of the perimeter (surface area in 3D) of the droplet.

*Sherwood number for a single droplet*

The Sherwood number($Sh$) for a droplet is defined as(Crowe et al., 2011):

$$Sh = \frac{\dot{m}"D}{\rho_{ps}D_v(Y_s - Y_\infty)} \tag{39}$$

where $D_v$ is the diffusion coefficient of Al vapor in air at the droplet surface. $\dot{m}"$ is the average mass-flux per unit area at the droplet surface. $\dot{m}"$ is obtained by taking the average of local mass-flux at droplet surface($\dot{\omega}"$) using the following equation:

$$\dot{m}" = \frac{\oint_S \dot{\omega}" dl}{S} \tag{40}$$

$Y_\infty$ in eq. (39) is the mass-fraction of Al vapor in the free-stream flow far upstream of the droplet. In our current calculations, the incoming free-stream flow does not contain any Al vapor. Therefore, $Y_\infty = 0$ in the current calculations. $Y_s$ in eq. (39) is the mass-fraction of Al vapor in a saturated air-vapor mixture at the temperature $T_p$. $Y_s$ is computed from the following relation:

$$Y_s = \frac{Mw_{Al} p_{sat}}{Mw_{air}(p_{ps} - p_{sat}) + Mw_{Al} p_{sat}} \qquad (41)$$

where $p_{sat}$ and $p_{ps}$ are the saturation vapor pressure of Al vapor at $T_p$ and the pressure of air behind the shock. $Mw_{Al}$ and $Mw_{air}$ are the molecular weight of Al and air.

In the current calculations, the $Sh$ varies with time as the shock wave travels past the droplet. The time-averaged $Sh$ and $Nu$ are computed as follows:

$$\overline{Nu} = \frac{\int_{t_1}^{t_2} Nu(\tau) d\tau}{t_2 - t_1} \qquad (42)$$

$$\overline{Sh} = \frac{\int_{t_1}^{t_2} Sh(\tau) d\tau}{t_2 - t_1} \qquad (43)$$

where $t_1$ and $t_2$ are the lower and upper bound of the time-averaging window. The time-averaged Sherwood number, $\overline{Sh}$ and $\overline{Nu}$ are the representative non-dimensional quantities used to characterize the vaporization and the heat-transfer rate of a droplet for a given $M_s$ and $Re_D$.

## 2.6 Construction of Surrogate Models for Sherwood number

The numerical method described above is used to calculate $\overline{Sh}$ and $\overline{Nu}$ from resolved numerical simulations of shock-droplet interaction at several locations within the parameter-space of $1.1 \leq M_s \leq 3.5$ and $100 \leq Re_D \leq 1000$. Following that, the simulation-based data is used to generate surrogate models for $\overline{Nu}$ and $\overline{Sh}$ a function of $M_s$ and $Re_D$, using the Modified Bayesian Kriging (MBKG) (Das et al., 2018b, 2018a; Gaul, 2014b; Sen et al., 2015, 2017a) method. The surrogate model serves as a bridge between the meso- and macro-scale. It also provides insights into the variation of vaporization and heat transfer with the flow parameters $M_s$ and $Re_D$.

In the current study, an initial surrogate is constructed using 16 mesoscale computations. The MBKG adaptive sampling strategy(Das et al., 2018b; Sen et al., 2017b) is then used to select inputs for the subsequent mesoscale simulations by adding 16 inputs each time. The final surrogate is created using 64 mesoscale computations. The locations of the mesoscale computations are shown in Figure 2.

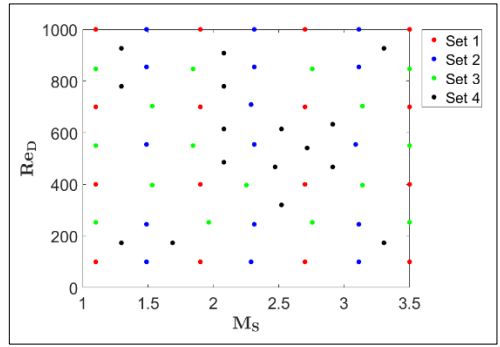

Figure 2 Locations of the meso-scale simulations in the parameter-space

## 3    Results and discussion

The numerical framework discussed in Section 2.1-2.5 is used to study Al droplet vaporization at several $M_s$ and $Re_D$. The current numerical solver has been extensively tested and validated against several benchmark problems in previous studies (Das et al., 2018b, 2018a, 2017). In this work, the current numerical framework is validated against an exact solution for a 1D air-liquid Al shock tube problem. The vaporization and heat-transfer from a cylindrical droplet during interaction with incoming shock waves are studied through several simulations. Following that, a simulation-driven model for $\overline{Sh}$ is developed as a function of $M_s$ and $Re_D$.

### 3.1   1D air-liquid Al shock-tube problem

A 1D air-liquid Al shock-tube problem is solved to validate the current hyperbolic equation solver coupled with the levelset based sharp-interface method and the RS-GFM. Figure 3 shows the schematic of the initial setup of this problem. For this problem, a 1m long shock-tube is considered. Initially, the air-liquid Al interface is assumed to be located at 0.5m from the left end of the shock-tube. Air at a low pressure of $10^5$Pa assumed on the left-side of the interface and pressurized liquid Al at $10^8$Pa is on the right side. The initial conditions in air and liquid Al are enumerated in the following table:

|  | Air | Liquid Al |
| --- | --- | --- |
| Pressure | $10^5 Pa$ | $10^8 Pa$ |
| Density | 1.0 kg/m³ | 2065.0 kg/m³ |
| Velocity | 0.0 m/s | 0.0 m/s |

Table 3 Initial conditions for the 1D air-liquid Al shock-tube problem

As the system is released to evolve from the above initial condition, the air-Al interface starts to move towards the left end of the shock-tube. A strong shock wave is generated in the air and a rarefaction wave is generated in the liquid Al at the interface. The shock travels through the air towards left-end of the shock-tube and a rarefaction wave travels through the liquid Al towards the right. The speed of the shock wave, the rarefaction waves and the contact discontinuity propagating through the materials are calculated from the exact solution of the Riemann problem(Toro, 2013) and compared with the numerical solution.

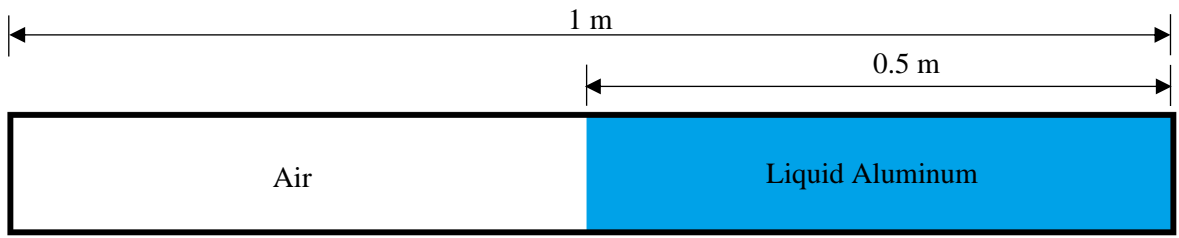

Initial conditions:

$P_{air} = 10^5 Pa$          $P_{Al} = 10^8 Pa$

$\rho_{air} = 1.0 \ kg/m^3$          $\rho_{Al} = 2065.0 \ kg/m^3$

$u_{air} = 0.0 \ m/s$          $u_{Al} = 0.0 \ m/s$

Figure 3 The initial conditions for the 1D air-liquid Al shock tube problem

The exact solution of the current Riemann problem is compared with the numerical solution in Figure 4. In the numerical solution of this problem, the interface is represented through the levelset based sharp-interface method. The flow-variables in the air and the liquid Al are coupled through the Riemann solver based GFM. The pressure, density and velocity distribution across the length of the shock tube obtained from the numerical solution at t=100μs are compared with the exact solution in Figure 4(a), (b) and (c) respectively. Results obtained from four different grid resolutions corresponding to the grid sizes of 5mm, 2.5mm, 1.25mm and 0.625mm are shown in Figure 4.

Figure 4 shows that the solution state predicted by the current solver agrees well with the exact solution. Figure 4 (a) shows that the location of the shock wave in air at $x = 0.46$m and the location of the rarefaction wave in liquid Al at $x = 0.87$m in the numerical result agree with the exact solution. The intermediate pressure($\sim 1.049 \times 10^5$Pa) between the shock wave and the rarefaction wave across the interface is predicted accurately using the current numerical method. The density distribution in Figure 4 (b) shows that the location of the contact discontinuity at the gas-liquid interface predicted by the current sharp-interface method agrees with the exact solution. Figure 4(c) shows that the current prediction of the velocity distribution across the shock-tube is also in good agreement with the exact solution. The smearing of the solution near the shock wave and the rarefaction wave due to numerical diffusion is seen in the results obtained from the coarse mesh simulations in Figure 4. However, the thickness of the shock and the rarefaction wave decreases as the grid resolution is increased.

For an error analysis, the $L_1$ error$(\epsilon_p(t))$ in the solution of the pressure field is computed as follows:

$$\epsilon_p = \frac{\int |p_{exact} - p_{num.}| dx}{\int |p_{exact}| dx} \tag{44}$$

where $p_{exact}$ and $p_{num.}$ are the exact solution and the numerical solution for the pressure field for a given grid resolution. $\epsilon_p$ of the numerical solutions is given in the following table:

| Grid size($\Delta x$) | $\epsilon_p$ |
|---|---|
| 0.005m | 0.145 |
| 0.0025m | 0.086 |
| 0.00125m | 0.048 |
| 0.000625m | 0.026 |

Table 4 The error in the numerical solution of the 1-D air-liquid aluminum shock tube problem at $100\mu s$ obtained from different grid-resolutions.

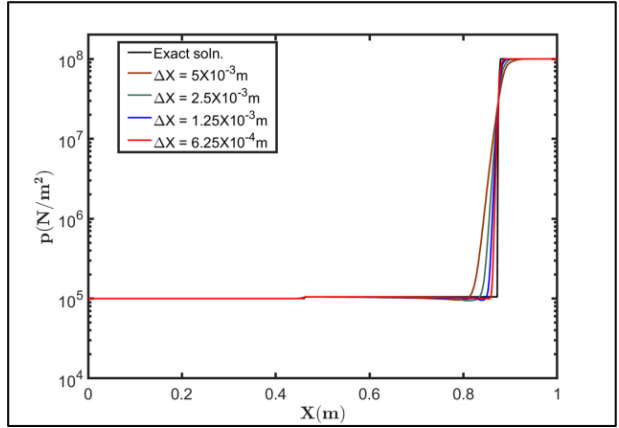

a)

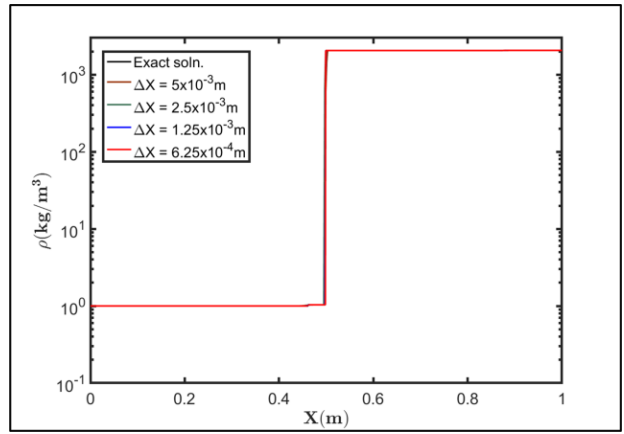

b)

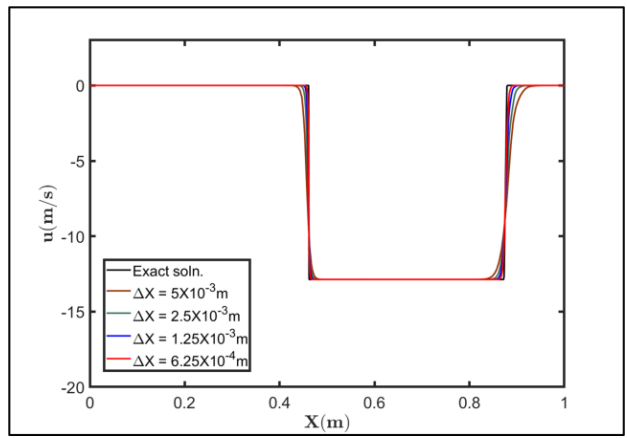

c)

Figure 4 Comparison of the (a) pressure (b) density and (c) velocity obtained from the analytical solution and the simulations result of the air-liquid Al shock tube problem after 100μs.

Table 4 shows that $\epsilon_p$ monotonically decreases with the grid-size. The results in Figure 4 and the error analysis in Table 4 delineate that the solution obtained using the current numerical framework converge to the exact solution as the grid resolution increased. Furthermore, the results in presented in Figure 4 show that the current solver predicts wave propagation through the air and the liquid Al accurately. Therefore, the flow variables in the gas and the liquid phase at the interface are accurately coupled using the current RS-GFM.

## 3.2 Mach 3.5 shock interaction with a cylindrical Al droplet

Vaporization of a cylindrical Al droplet under shock loading is studied to demonstrate the physics involved at the different stages of shock-droplet interaction. For this study, a Mach 3.5 shock interaction with a cylindrical Al droplet of diameter $D$=3.84 μm is considered. The initial set-up of the computational domain is shown in Figure 5. Neumann boundary conditions are applied at the boundaries of the computational domain. The initial conditions used in this study are enumerated in the following table:

|  | $\rho(\text{kg/m}^3)$ | $p(\text{Pa})$ | $u(\text{m/s})$ | $T(\text{K})$ |
|---|---|---|---|---|
| **Pre-shocked air $\left(\frac{X}{D} \geq 2.25\right)$** | 1.204 | 101325.0 | 0.0 | 293.0 |
| **Post-shocked air $\left(\frac{X}{D} < 2.25\right)$** | 5.131 | 1431215.625 | 1201.207 | 971.81 |
| **Droplet** | 2003.0 | 376120.7565 | 0.0 | 2743.0 |

Table 5 Initial conditions for the simulation of Mach 3.5 interaction with a cylindrical Al droplet od diameter 3.84 μm.

The $Re_D$ and $We_D$ calculated based on the post-shock conditions are 1000.0 and 31.56 respectively. The Diameter of the droplet ($D$) is selected as the reference length-scale in this problem. The reference timescale ($\tau$) of the problem is defined as the shock passage time across the droplet diameter $\left(\frac{D}{u_s}\right)$. The reference pressure($p_0$) is selected as 1 bar.

Initially, five different grid resolutions corresponding to 50, 100, 150, 200, and 250 grid points across the droplet diameter are selected for a grid convergence study. The $\dot{m}''$ computed using these different grid resolutions are plotted against the nondimensionalized time $t^*\left(=\frac{t}{\tau}\right)$ in Figure 6(a). Figure 6(a) shows that the calculated $\dot{m}"$ converges with grid refinement.

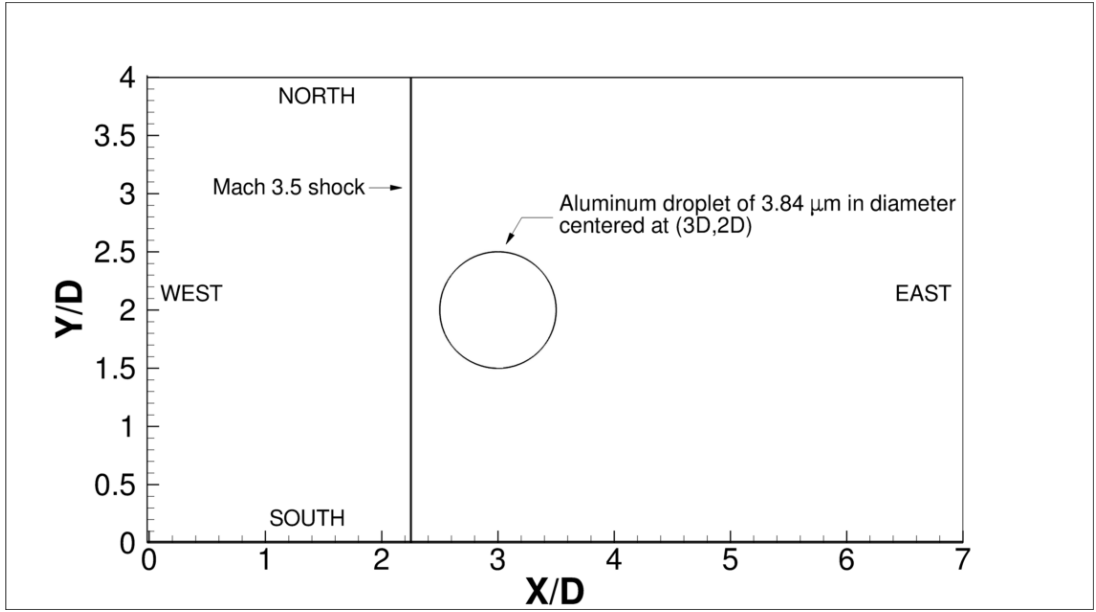

Figure 5 Initial set-up of the computational domain for the simulation of shock($M_s = 3.5$) interaction with an evaporating droplet($Re_D = 1000$).

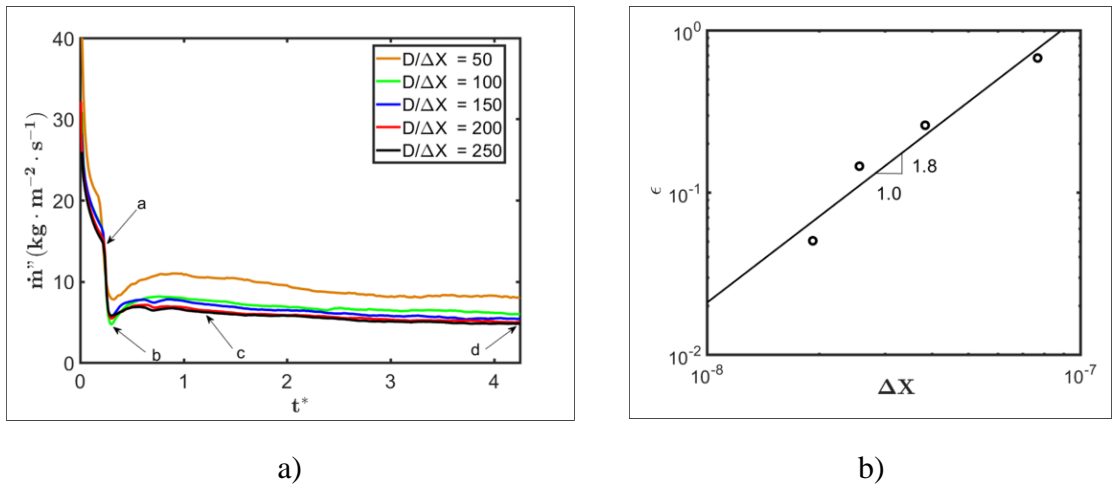

a)  b)

Figure 6 The grid-convergence study for Mach 3.5 shock interaction with a 3.84μm Al droplet. a) The time-evolution of the $\dot{m}"$ during the shock-droplet interaction. b) The error in the calculation of $\bar{\bar{m}}$ using different grid-sizes.

The time-averaged vaporization mass flux ($\overline{\dot{m}''}$) at the surface of the Al droplet during interaction with the incoming shock wave is computed from:

$$\overline{\dot{m}''} = \frac{\int_{t_1^*}^{t_2^*} \dot{m}'' dt}{t_2^* - t_1^*} \tag{45}$$

where, $t_1^* = 1.25$ and $t_2^* = 4.25$. The error($\epsilon$) in the calculation of $\overline{\dot{m}''}$ is evaluated with respect to the results obtained from the finest grid as follows:

$$\epsilon = \left| \frac{(\overline{\dot{m}''} - \overline{\dot{m}''}_{ref})}{\overline{\dot{m}''}_{ref}} \right| \tag{46}$$

where $\overline{\dot{m}''}_{ref}$ is obtained using $\frac{D}{\Delta x} = 250$. The $\overline{\dot{m}''}$ and $\epsilon$ computed from the different grid resolutions are tabulated below:

| Grid | Resolution ($\frac{D}{\Delta x}$) | $\overline{\dot{m}''}$ (kg·m$^{-2}$·s$^{-1}$) | $\epsilon$ |
|---|---|---|---|
| GRID1 | 50 | 8.78 | 0.68 |
| GRID2 | 100 | 6.61 | 0.26 |
| GRID3 | 150 | 6.01 | 0.15 |
| GRID4 | 200 | 5.52 | 0.05 |
| GRID5 | 250 | 5.25 | - |

Table 6 $\overline{\dot{m}''}$ computed from numerical calculations of $M_s = 3.5$ shock interaction with an Al droplet of $3.84\mu m$ in diameter using different grid resolution

The values of $\overline{\dot{m}''}$ in Table 6 show that the vaporization rate of the Al droplet computed using the current method converges as the grid resolution of the calculation is increased. $\epsilon$ is plotted against grid resolution in Figure 6(b), showing that the error $\epsilon$ decreases monotonically with grid refinement. Figure 6(a) shows that the $\dot{m}''$ computed from the GRID4 and GRID5 are almost identical. $\overline{\dot{m}''}$ computed from GRID4, shown in Table 6, has negligible error with respect to GRID5. Only 5% more accuracy is gained by using GRID5 instead GRID4. However, GRID5 is 56% more computationally expensive than GRID4. To balance computational cost and accuracy, GRID4 is used in the rest of the simulation to develop the simulation-derived surrogate model for $Sh$ from the simulations.

Figure 6(a) shows the transient unsteadiness in $\dot{m}''$ of the Al droplet during the interaction with the shock-wave. To further understand the reason for the unsteadiness in the $\dot{m}''$, the contours of numerical Schlieren, pressure and Al-vapor mass fraction at four different time instances are plotted in Figure 7. The markers "a", "b", "c" and "d" in Figure 6(a) represent the corresponding time instances of the contour plots shown in Figure 7 (a), (b), (c), and (d) respectively.

The current calculation is initiated with a heated Al droplet in quiescent air at STP (standard temperature and pressure). The incident Mach 3.5 shock-wave is placed *0.25D* away from the droplet surface. Initially, the Al droplet is at its boiling point (2743.0K). Since the surrounding air does not contain any Al vapor at the beginning of the simulation, the Al vapor pressure around the droplet is zero. Under these conditions the Al droplet starts to vaporize at a high-rate immediately after the calculations are started. This can be seen in Figure 6(a). Figure 6(a) shows that the $\dot{m}''$ is highest (~40 kg.m$^{-2}$.s$^{-1}$) at the beginning of the simulation. The $\dot{m}$ decreases rapidly between $t^* = 0 - 0.25$ as Al vapor accumulates around the droplet. This accumulation occurs because, during this period, the shock wave has not yet reached the droplet surface and the droplet is immersed in the quiescent air. The accumulation of Al vapor increases the vapor pressure at the surface of the droplet. The rapid increase in vapor-pressure thereafter suppresses the $\dot{m}''$, as seen in Figure 6(a).

Figure 7(a) shows the contour plots of numerical Schlieren, pressure($p$) and Al vapor mass fraction($Y_{Al}$) shortly after the incident shock wave has interacted with the droplet surface. The contours of $Y_{Al}$ in Figure 7(a) show the layer of Al vapor accumulated around the droplet. The initial high rate of vaporization and heat transfer increase the temperature and the pressure of the quiescent air surrounding the droplet. The rapid increase in the temperature and pressure of the air around the droplet initiates a thermally-induced acoustic wave. The thermally-induced acoustic wave can be seen in the numerical Schlieren in Figure 7(a).

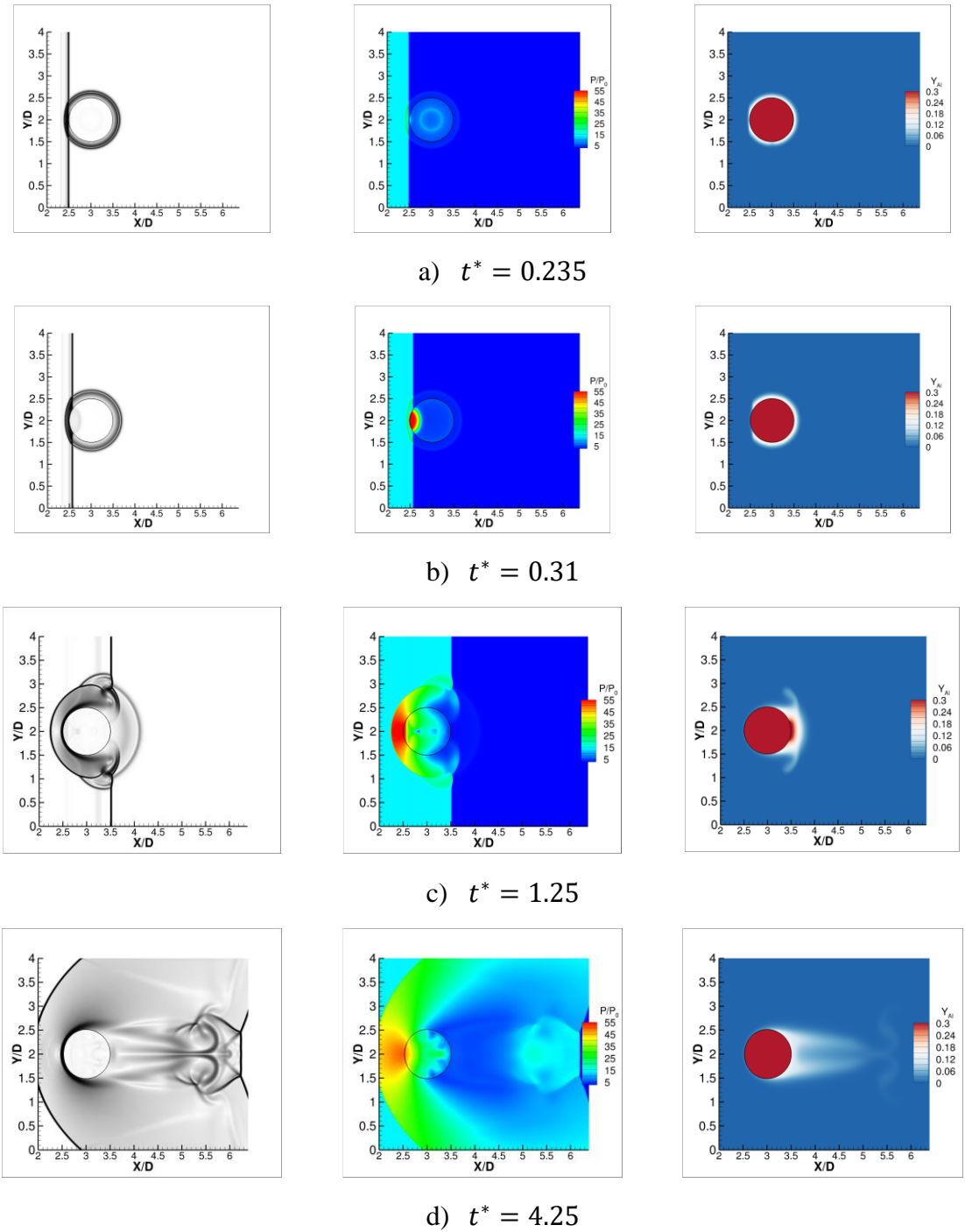

Figure 7 Sequence of Schlieren (left column), pressure(middle column) and contours of $Y_{Al}$ (right column) obtained from numerical simulation of Mach 3.5 shock interaction with an Al droplet($Re_D = 1000$).

The outward traveling thermally-induced acoustic wave initiated at the droplet surface also induces another acoustic wave that travels radially inside the droplet. The induced acoustic wave inside the droplet is seen in the pressure contour plotted in Figure 7 (a). The higher pressure inside the droplet is due to the surface tension at the liquid Al-air interface.

The interaction of the incident shock with the air-liquid-Al interface of the droplet induces a reflected wave in the air and a transmitted wave in the liquid Al. The reflected and transmitted shock waves are seen in Figure 7(b). Figure 7(b) also shows that the droplet surface is pressurized as the shock-wave impinges on the droplet surface. The shock-induced pressurization causes the vapor-pressure to increase abruptly at the droplet surface and the $\dot{m}''$ decreases further. Figure 6(a) shows this behavior between $t^* = 0.2335 - 0.31$ and the $\dot{m}''$ attains its lowest value at $t^* = 0.31$.

In contrast to the case shown in Figure 7, Figure 6 (a) shows that $\dot{m}''$ increases as the shock wave travels farther over the droplet. $\dot{m}''$ increases because the vapor-pressure around the droplet decreases after $t^* = 0.31$. The numerical Schlieren, $p$ and $Y_{Al}$ contour in Figure 7(c) show that as the shock wave travels past the droplet the layer of Al vapor accumulated is stripped off the droplet surface by the high-speed flow. This is accompanied by the depressurization at the leeward side of the droplet caused by expansion waves. As a cumulative effect, the vapor pressure around the droplet decreases and the $\dot{m}''$ increases.

As the flow evolves further, a wake is formed behind the droplet. Figure 7(d) shows that the evaporated Al remains within the boundary-layer and is carried into the wake of the Al droplet. After the initial unsteadiness, the vaporization rate stabilizes as a quasi-steady boundary-layer is formed around the droplet. The stabilization of the $\dot{m}''$ is also observed in Figure 6(a) after the shock wave has passed over the droplet. A time-average of the quasi-steady $\dot{m}''$ between $t_1^* = 1.25$ and $t_2^* = 4.25$ is considered as the representative quasi-steady vaporization rate of a droplet for a given $M_s$ and $Re_D$. In the following sections, the effects of $M_s$ and $Re_D$ on the vaporization rate of a droplet is studied.

## 3.3 The effect of $M_s$ and $Re_D$ on the vaporization and heat-transfer from the Al droplet

The $\dot{m}"$ and $\dot{q}"$ are influenced by the flow-field around the droplet. The axial-velocity$\left(\frac{u}{U_0}\right)$, pressure$\left(\frac{p}{p_0}\right)$, vapor mass-fraction$(Y_{Al})$, temperature$\left(\frac{T}{T_0}\right)$ and the vapor pressure$(p_v)$ field around the droplet under different flow conditions are compared in Figure 8 to study the influence of $M_s$ and $Re_D$ on the flow-field around the droplet. The flow-fields computed from three different cases, e.g., $M_s = 3.5, Re_D = 100$; $M_s = 3.5, Re_D = 1000$ and $M_s = 1.1, Re_D = 1000$ at $t^* = 4.25$ are discussed in the following subsections.

### *The effects of $Re_D$ on the flow-field around the droplet*

The flow-field computed for $Re_D = 100$ and 1000 are presented in the left and the middle column of Figure 8. $M_s$ is 3.5 for both cases. The comparison of the flow-fields in the left and the middle column of Figure 8 shows the effect of $Re_D$ on the flow-field when $M_s$ is kept fixed.

The contour plots of $\frac{u}{U_0}$ along with streamlines are shown in Figure 8(a). Figure 8(a) shows the development of the boundary layer over the droplets and shape of the recirculation region in the wake of the droplet under different flow conditions. The comparison of the plots in the left and the middle column of Figure 8(a) shows that the boundary layer at the droplet interface is thinner (when compared via the non-dimensional length scale) at $Re_D = 1000$ compared to $Re_D = 100$. The comparison of the streamlines in the left and the middle column of Figure 8(a) shows that the flow separation is expedited at $Re_D = 1000$ when compared to $Re_D = 100$. The early flow separation and the lower rate of diffusive transport of momentum results in a longer and wider recirculation bubble in the wake of the droplet at the higher $Re_D$. Effects of higher diffusive transport are found to dominate the flow-field at $Re_D = 100$.

The comparison of the contour plots in the left and the middle column of Figure 8(b) shows the differences in the pressure-field around the droplet at $Re_D = 100$ and $Re_D = 1000$. The pressure inside the droplet in the left column of Figure 8(b)($Re_D = 100$) is higher than the droplet in the middle column ($Re_D = 1000$).

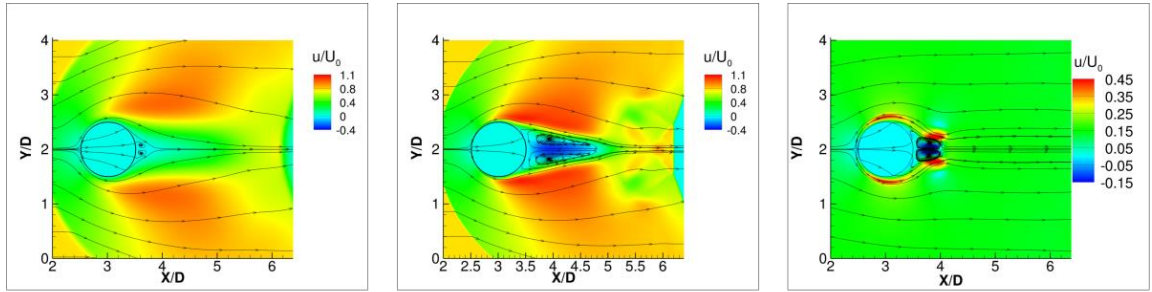

a)

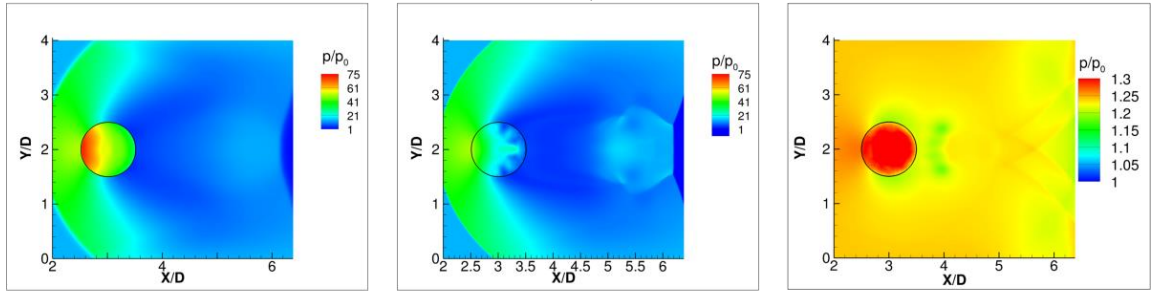

b)

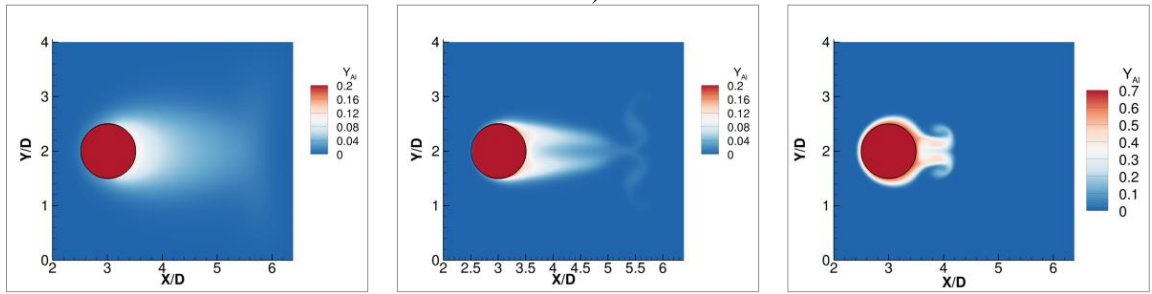

c)

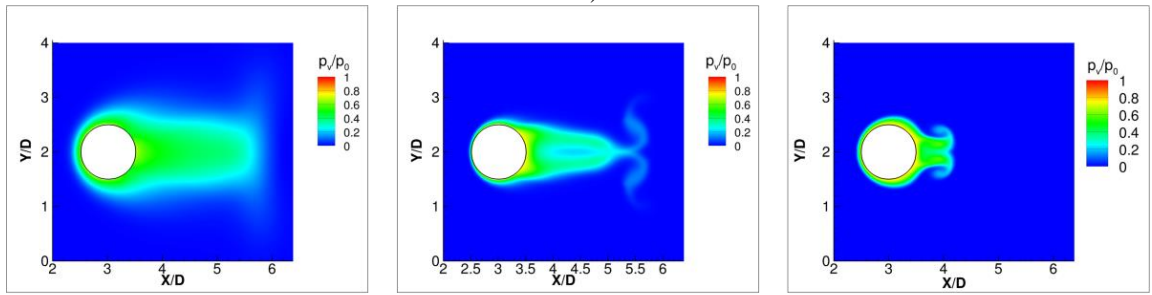

d)

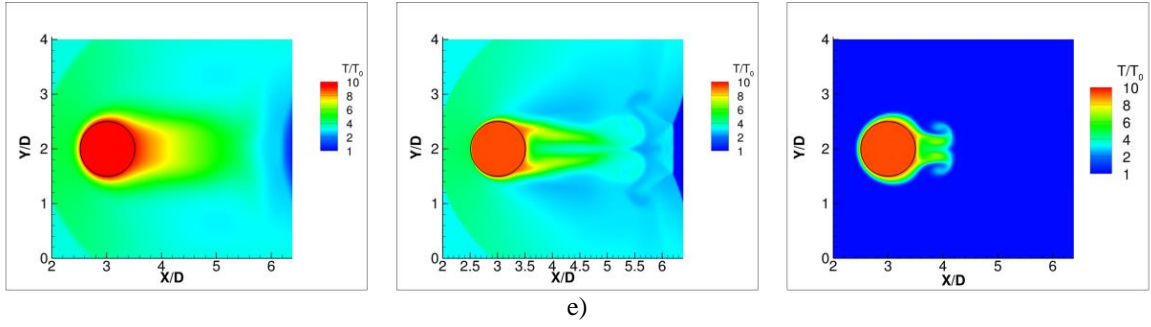

e)

Figure 8 Comparison of a) axial-velocity b) pressure c) vapor mass-fraction d) temperature and e) vapor-pressure distribution at $t^* = 4.25$ around the droplets under three different flow conditions. In the left column $M_s = 3.5, Re_D = 100$, in the middle column $M_s = 3.5, Re_D = 1000$ and in the right column $M_s = 1.1, Re_D = 1000$. In the above plots, $U_0 = shock\ speed$, $P_0 = 101325.0 Pa$ and $T_0 = 293.15\ K$

This is because the droplet in the left column of Figure 8 has a smaller diameter($D = 0.39\mu m$) compared to the droplet in the middle column ($D = 3.9\mu m$). The pronounced effect of surface tension in the smaller droplet increases the pressure within the droplet significantly. Apart from the difference in the pressure inside the droplets, the pressure field around the droplet is marginally different for $Re_D = 100$ and $Re_D = 1000$. In both cases, the leeward side of the droplets is exposed to a low-pressure wake region. However, the early separation at $Re_D = 1000$ causes the droplet in the middle column to have a larger area of the droplet surface exposed to the low-pressure wake than for the $Re_D = 100$ case shown in the left column.

The contour plots of $Y_{Al}$ in the left and the middle column of Figure 8(c) show that for the Mach number $M_s = 3.5$, the Al droplets vaporize at a higher rate from the leeward side because of the lower pressure. The vaporized Al accumulates in the wake of the droplet due to advection of the vapor contained in the boundary layer into the wake. But at the same time, mass-diffusion also transports the Al vapor away from the droplet. Since diffusive transport processes are more influential at smaller length-scales, the rate for diffusive transport is higher in the case of $Re_D = 100$, i.e. for a smaller droplet. The evaporated mass is transported away at a higher rate from the surface of smaller droplets and the amount of vapor accumulated around a smaller droplet surface is comparatively lower than around a bigger droplet when the $M_s$ is kept unchanged. This is observed in the contour plot of $Y_{Al}$ in Figure 8(c). Lower concentration of $Y_{Al}$ near the droplet surface is observed for $Re_D = 100$ compared to $Re_D = 1000$. As a result, $p_v$ around a droplet decreases with the decrease in $Re_D$. The contour plots of $\frac{p_v}{p_0}$ in the left and the middle column of Figure 8(d) demonstrate that the vapor pressure is lower around the droplet at $Re_D = 100$ compared to $Re_D = 1000$ for $M_s = 3.5$. The $p_v$ at the droplet surface controls the $\dot{\omega}"$. The $\dot{\omega}"$(Eq. (6)) increases with the decrease in $p_v$. Therefore, for constant $M_s$, $\dot{m}"$ increases with the decreasing $Re_D$.

The contours of normalized temperature $\frac{T}{T_0}$ during $M_s = 3.5$ shock interaction with a droplet at $Re_D = 100$ and $Re_D = 1000$ are compared in the left and the middle column of Figure 8(e). Figure 8(e) exhibits a thicker thermal boundary-layer over the smaller droplet at $Re_D = 100$ compared to the larger droplet at $Re_D = 1000$. As a consequence, the temperature around the smaller droplet, corresponding to $Re_D = 100$, is higher than the larger droplet with $Re_D = 1000$, as observed in Figure 8(e). The temperature of the air around the droplet surface has noticeable effects on the $\dot{\omega}"$(Eq. (6)). $\dot{\omega}"$ increases as the temperature of the gas-phase is increased. Therefore, the $\dot{m}"$ will be positively impacted by the increase in the heat conduction from the droplet to the surrounding air at the low Reynolds numbers.

### *The effects of $M_s$ on the flowfield around the droplet*

Results obtained from a calculation of shock-droplet interaction at $M_s = 1.1$ and $Re_D = 1000$ are presented in the rightmost column of Figure 8. The differences in the flow-field around the droplet during interaction with Mach 3.5 and Mach 1.1 shocks are shown by comparing the contour plots presented in the middle and the rightmost columns of Figure 8.

The comparison of the contours of $\frac{u}{U_0}$ in the middle and the right column of Figure 8(a) shows that the thickness of the boundary-layer over the droplets is comparable in the non-dimensional length scale for $Re_D = 1000$ in both calculations. However, the flowfield in the vicinity of the droplets at $M_s = 1.1$ and 3.5 are remarkably different. At $M_s = 1.1$, the post-shock flow Mach number is 0.154 compared to Mach 1.92 for the $M_s = 3.5$ case. The high-speed post-shock flow at $M_s = 3.5$ is rapidly decelerated across the shocks and accelerated across the expansion-waves around the droplet. Such features of a compressible flowfield are absent at $M_s = 1.1$. At $M_s = 1.1$, the flow behind the shock is essentially in the low Mach number incompressible regime. Figure 8(b) shows that the pressure around the droplet does not vary significantly during the interaction with a Mach 1.1 shock. The pressure distribution around the droplet influences the evaporation mass flux at the droplet surface. Figure 8(c) shows that, unlike during interaction with a Mach 3.5 shock, the droplet vaporizes at relatively more even rate from all sides as the Mach 1.1 shock travels across. Figure 8(c) also shows a higher accumulation of Al vapor around the droplet at the $M_s = 1.1$ compared to $M_s = 3.5$. This is because of the reduced rate of convective mass transport from the droplet surface at the lower Mach number. Therefore, the mass fraction $Y_{Al}$ in Figure 8(c) and subsequently $p_v$ in Figure 8(d), is higher around the droplet surface for $M_s = 1.1$ when compared to $M_s = 3.5$. As a result, $\dot{m}''$ increases when the $M_s$ is increased from 1.1 to 3.5.

Figure 8(e) shows the temperature contour around the droplet at $M_s = 1.1$. The comparison of the temperature contours in Figure 8(e) show a lower temperature in the gaseous phase surrounding the droplet at $M_s = 1.1$. This is because, the flow-field remains subsonic behind the Mach 1.1 shock, while the Mach 3.5 shock entrails a supersonic flow behind it. As a result, a reflected bow shock is formed around the droplet interacting with Mach 3.5 shock, while, a sonic acoustic wave reflects back from the droplet surface during the interaction with Mach 1.1 shock. The reflected bow-shock raises the temperature around the droplet significantly, which leads to a higher temperature around the droplet interacting with Mach 3.5 shock compared to the Mach 1.1 shock. The temperature in the boundary-layer is further raised at the higher flow-speed because of the viscous dissipation. At $M_s = 3.5$, there will be a higher gradient of axial-velocity in the boundary-layer around the droplet. This leads to a higher rate of heating in the boundary-layer due to viscous dissipation. The viscous effect also adds on to increase the overall temperature around the droplet surface at $M_s = 3.5$ compared to $M_s = 1.1$. The increase in the temperature of the gaseous phase around the droplet increases the $\dot{m}''$.

The difference in the flow-fields around the droplets interacting with shocks under different flow conditions play a role in modulating the vaporization mass-flux($\dot{m}''$) and heat-flux ($\dot{q}''$) from the droplets. The $\dot{m}''$ and $\dot{q}''$ of the droplets under different flow conditions are compared in the following section.

### *The effect of $Re_D$ on $Sh$ and $Nu$ of a vaporizing Al droplet:*

The effect of $Re_D$ on the vaporization of Al droplets interacting with a Mach 3.5 shock is studied by comparing the $\dot{m}''$ and $Sh$ for different droplet sizes. Droplets of different sizes are studied, corresponding to $Re_D$ = 100, 400, 700, and 1000. The $\dot{m}''$ computed from the simulations are plotted in Figure 9(a). Figure 9(a) shows that the $\dot{m}''$ decreases with the increase in $Re_D$. The $\overline{\dot{m}''}$ decreases from 11.63 kg/m²s to 5.45 kg/m²s as the $Re_D$ is increased from 100 to 1000. As shown in the previous section, the rate of the diffusive transport of Al vapor from the droplet surface decreases with the increase in $Re_D$ which leads to accumulation of the Al vapor around the droplet at the higher $Re_D$. The accumulation of the Al vapor increases the vapor pressure($p_v$) at the droplet surface as shown previously in Figure 8(d). Furthermore, the temperature within the boundary layer around the droplet also decreases marginally with increase in $Re_D$. The cumulative effect of the increase in $p_v$ and the decrease in $T$ around the droplet surface at the higher $Re_D$ suppresses the vaporization rate $\dot{m}''$ of the droplet.

On the contrary, Figure 9(b) shows that the $Sh$ of a droplet for $M_s = 3.5$ increases with $Re_D$. The mass-flux due to the diffusive transport is estimated as $\rho_{ps} D_v \left(\frac{Y_s - Y_\infty}{D}\right)$ (Eq. (39)). Therefore, the rate of diffusive transport of the vapor from the droplet surface decreases when the size of the droplet and consequently the $Re_D$ is increased. However, $\rho_{ps} D_v \left(\frac{Y_s - Y_\infty}{D}\right)$ decreases at a higher rate than $\dot{m}''$ when the droplet size is increased. As a result, $Sh$ increases with $Re_D$ when $M_s$ is kept fixed.

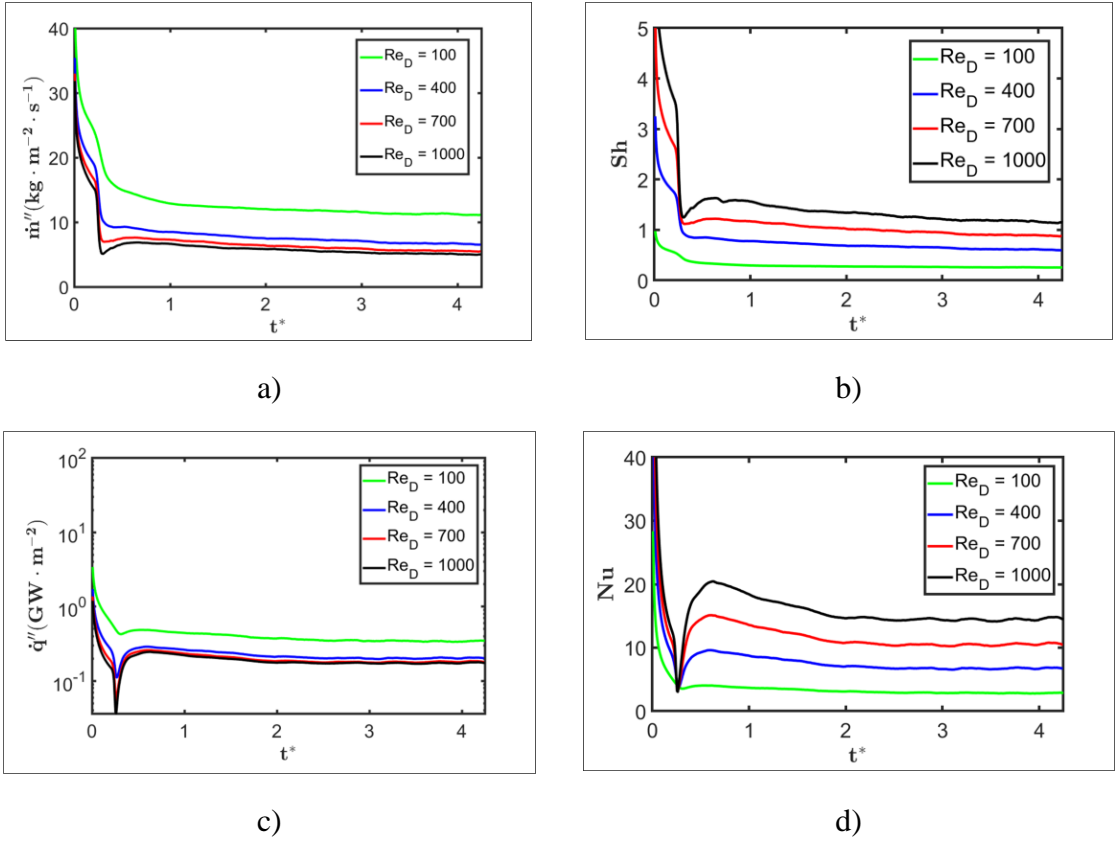

Figure 9 The comparison of a) $\dot{m}''$, b) $Sh$, c) $\dot{q}''$ ,and d) c) $Nu$ of Al droplet during interaction with a $M_s = 3.5$ shock-wave at $Re_D = 100, 400, 700$ and $1000$.

The effect of $Re_D$ on the heat transfer from the droplet is studied next. The $\dot{q}''$ and $Nu$ for $M_s = 3.5$ and $Re_D = 100, 400, 700$ and $1000$ are compared in Figure 9(c) and (d), respectively. Figure 9 (c) shows that the $\dot{q}''$ decreases as $Re_D$ is increased. $\dot{q}''$ is higher at the smaller $Re_D$ because the thermal conduction is dominant at the smaller length scales. As the droplet size and thus the $Re_D$ is increased, the thickness of the thermal boundary layer on the droplet also increases. With the increase in the thickness of the boundary layer, the local temperature gradient at the droplet surface decreases. Therefore, $\dot{q}''$ decreases with increasing $Re_D$. However, Figure 9(d) shows that the $Nu$ increases with $Re_D$ or the droplet size. The $Nu$ represents the ratio of the convective and the diffusive transport of the heat from the droplet surface. The average heat-flux at the droplet surface due to thermal diffusion, i.e. $k_{air}\left(\frac{T_{ps}-T_\infty}{D}\right)$ (see Eq.(37)) decreases as the droplet size is increased. However, $k_{air}\left(\frac{T_{ps}-T_\infty}{D}\right)$ decreases faster than $\dot{q}"$ as $Re_D$ is increased. Consequently, $Nu$ increases with the increase in $Re_D$.

### *The effect of $M_s$ on the $Sh$ and $Nu$ of an Al droplet:*

The effect of $M_s$ on the $Sh$ and $Nu$ of an Al droplet is studied by comparing the results obtained for $M_s = 1.1, 1.9, 2.7$ and $3.5$ in Figure 10. The $Re_D$ is kept fixed at 1000 for these calculations. To keep the $Re_D$ constant, $D$ is decreased from 236µm to 3.9µm as the $M_s$ is increased from 1.1 to 3.5.

Figure 10 (a) shows that the $\dot{m}"$ increases with $M_s$. $\overline{\dot{m}"}$ increases from 1.2 kg/m²s to 5.45 kg/m²s as the shock strength is increased from $M_s = 1.1$ to $M_s = 3.5$. As discussed earlier, the shock-induced compression of the gaseous phases at different $M_s$ influences the $p_v$, $T$ and the transport of the Al vapor around the droplet significantly. The strength and the locations of the shocks and the expansion waves around the droplet changes as the $M_s$ is increased from 1.1 to 3.5 and post-shock flow changes from subsonic to a supersonic flow. The cumulative effect of such features of compressible flows around the droplet causes the $\dot{m}"$ to increase with $M_s$.

The $Sh$ of the droplet at $M_s = 1.1, 1.9, 2.7$ and $3.5$ computed from the current simulations are compared in Figure 10(b). The $Sh$ remains comparable for $M_s = 1.1$ and 1.9. However, $Sh$ increases when $M_s$ is increased further. Such behavior of the $Sh$ is due to the effects of the shock-induced compression of the gaseous phase around the droplets at different $M_s$ on the rate of the diffusive mass-transport from the droplet surface, $\rho_{ps}D_v\left(\frac{Y_s-Y_\infty}{D}\right)$. The compression of the gaseous phase $\left(\frac{\rho_{ps}}{\rho_{us}}\right)$ around the droplet changes drastically as $M_s$ increases from 1.1 to 1.9 and the post-shock flow makes a transition from subsonic to supersonic speed. Furthermore, $D_v$ and $\rho_{ps}$ increases with $M_s$, while, $Y_s$ decreases. As a result, $\rho_{ps}D_v\left(\frac{Y_s-Y_\infty}{D}\right)$ varies non-monotonically with $M_s$. Consecutively, $Sh$ varies non-monotonically with $M_s$.

The compression of the gaseous phase around the droplet at different $M_s$ influences the heat transfer from the droplet. The $\dot{q}"$ and $Nu$ computed for $M_s = 1.1, 1.9, 2.7$ and $3.5$ are compared in Figure 10(c). Figure 10(c) shows that $\dot{q}"$ increases with $M_s$. $\dot{q}"$ is dependent on the local temperature gradient at the droplet surface. The temperature gradient in the thermal-boundary layer around the droplet increases as the flow-velocity around the droplet increases and the droplet size decreases with the increasing $M_s$ at a fixed $Re_D$. As a result, $\dot{q}"$ increases with the $M_s$.

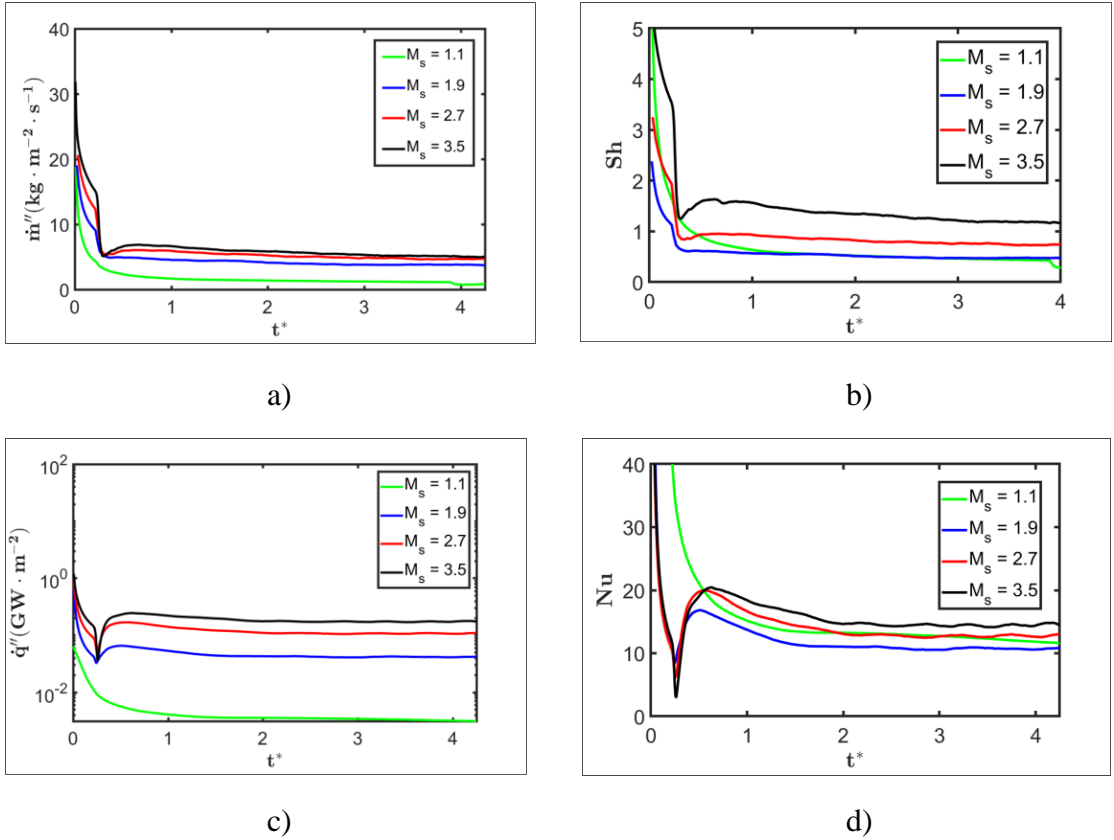

Figure 10 The comparison of a) $\dot{m}"$, b) $Sh$, c) $\dot{q}"$ ,and d) $Nu$ of the Al droplet during interaction with shock-waves of $M_s = 1.1, 1.9, 2.7$ and $3.5$. The $Re_D$ is 1000.

Figure 10 (d) shows that the $Nu$, similar to the $Sh$, varies non-monotonically with the $M_s$. The $Nu$ decreases in the beginning as $M_s$ is increased from 1.1 to 1.9. However, $Nu$ increases when $M_s$ is increased further. The rate of diffusive transport of heat at the droplet surface is heavily influenced by the compression of the gas-phase at the different $M_s$. As a result, the trend of $Nu$ with $M_s$ changes as the post-shock becomes supersonic from subsonic around $M_s = 1.9$. A further discussion on the effects of $M_s$ on $Nu$ can be found in (Das et al., n.d.).

## 3.4 Metamodel for $\overline{Sh}$ and $\overline{Nu}$ of a vaporizing Al droplet as a function of $M_s$ and $Re_D$

The data obtained from the simulations of shock interaction with cylindrical droplets is used to develop a surrogate model for the $\overline{Sh}$ and $\overline{Nu}$ of droplets in a shocked flow. The surrogate models are developed as a function of $M_s$ and $Re_D$ in the parameter space of $1.1 \leq M_s \leq 3.5$ and $100 \leq Re_D \leq 1000$. The MBKG method(Sen et al., 2017a) is used to develop the surrogate model from the simulation data. The location of the 64 simulations in the parameter space is shown in Figure 11(a). The surrogate model of $\overline{Sh}$ and $\overline{Nu}$ are shown in Figure 11(b) and (c), respectively.

The current surrogate model in Figure 11(b) shows that $\overline{Sh} < 1$ for the majority of the locations within the current parameter space. $\overline{Sh} > 1$ occurs only at high Mach numbers and Reynolds numbers, i.e. for $M_s > 3.13$ and $Re_D > 745$. This indicates that the diffusion mass-flux at the droplet surface is higher than the vaporization mass-flux in a substantial part of the current parameter-space. Therefore, Al vapor is transported away from droplet surface at a higher rate than the rate at which vaporization of the droplet occurs. The vaporization rate of the droplet is not limited by the rate of transport of the vapor from the droplet surface within the range of $M_s$ and $Re_D$ used in the current simulations.

The surrogate models in Figure 11(b) and (c) show that both $\overline{Sh}$ and $\overline{Nu}$ increase with $Re_D$. The rate of diffusive transport of the Al vapor and thermal energy from the droplet surface decreases as $Re_D$ increases; therefore, $\overline{Sh}$ and $\overline{Nu}$ monotonically increase with $Re_D$. On the other hand, the trends of $\overline{Sh}$ and $\overline{Nu}$ with respect to $M_s$ are non-monotonic over the parameter space. Initially the $\overline{Sh}$ and $\overline{Nu}$ decreases with the increase in the shock strength range of $M_s \sim 1.1 - 1.43$. However, $\overline{Sh}$ and $\overline{Nu}$ increase when the $M_s$ is increased further. Such non-monotonic behavior is caused by the compression of the flow around the droplet under different shock-loading as the post-shock flow Mach number varies from sub-sonic to super-sonic speeds within the parameter-space.

The current results in Figure 11(b) and(c) show that $\overline{Sh}$ is more sensitive to the change in $M_s$ than $\overline{Nu}$. The vaporization of the droplet is a pressure-driven phenomenon. The local vaporization mass-flux $\dot{\omega}''$ (Eq.(6)) is strongly influenced by the local vapor-pressure($p_v$) and temperature($T_s$) in the gaseous phase at the interface. $p_v$ at the droplet surface vary with $M_s$. As a result, $\overline{Sh}$ is a strong function of $M_s$.

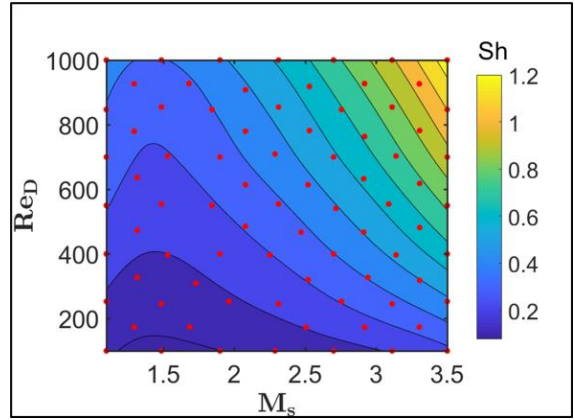

a)

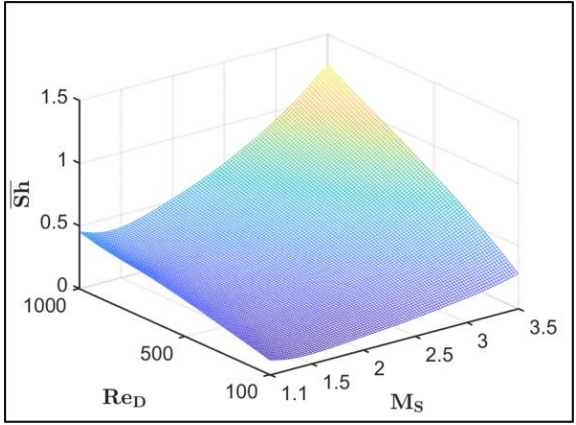

b)

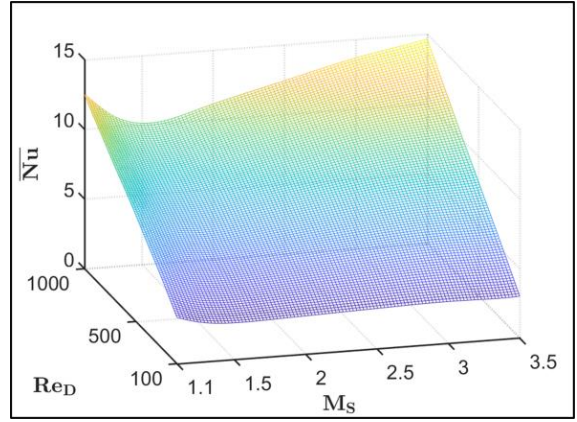

c)

Figure 11 Surrogate models of $\overline{Sh}$ and $\overline{Nu}$ obtained from the simulation-based data. a) 2D contour plot of the surrogate model for $\overline{Sh}$ along with the training data-points (marked in red). b) The shape of the surrogate model for $\overline{Sh}$. c) The shape of the surrogate model for $\overline{Nu}$.

On the other hand, $\overline{Nu}$ is relatively less sensitive to $M_s$. The heat-transfer between the droplet and the gas is governed by the rate of the convective and diffusive transport of thermal energy. The heat-transfer from the liquid droplet is influenced by the $Re_D$. The changes in the pressure and temperature distribution around the droplet marginally affects the $\overline{Nu}$. As a result, the $\overline{Nu}$ of a droplet is found to be more sensitive to the $Re_D$ than $M_s$.

Nevertheless, $\overline{Sh}$ and $\overline{Nu}$ both are affected by the change in $M_s$ in a shocked flow-field because of the compression of the gaseous phase around the droplet. As opposed to the previous models(Abramzon and Sirignano, 1989; Kreith et al., 2012; Ranz and Marshal, 1952) the current models take the effects of the compressibility of the flow into account by casting $\overline{Sh}$ and $\overline{Nu}$ as a function of $M_s$ and $Re_D$.

# 4 Conclusions

The physics of the vaporization of aluminum droplets in shocked-conditions is numerically investigated and models for the Sherwood number and the Nusselt number cast as functions of the shock Mach number and the Reynolds number are developed from the simulation-based data. A levelset-based sharp-interface method is used to perform simulations of shock interaction with vaporizing droplets at the various shock Mach numbers and Reynolds numbers. In this study, a kinetic model (Houim, 2011) of vaporization of the liquid droplets is used to calculate the local vaporization rates at the droplet surface. A modified Riemann solver based ghost fluid method is used to allow the characteristic waves to travel between the gaseous and the liquid phases separated by the sharp-interface. The interfacial jump conditions due to the surface tension and phase change at the interface are treated through an interfacial Riemann solver based ghost fluid method. The current sharp-interface method is used to perform calculations of shock-droplet interaction to study the effects of flow conditions -- spanning the parameter space characterized by the shock Mach number and the Reynolds number -- on the vaporization rate and the heat-transfer from the droplets.

The Sherwood number and the Nusselt number of the droplets subjected to different shock Mach number and the Reynolds number are computed from the current simulations. The results show that the the Sherwood number and the Nusselt number of the droplet increase monotonically with the Reynolds number. Diffusive transport becomes less significant than convective transport of the vapor and heat from the droplet surface as the Reynolds number increases. As a result, the Sherwood number and the Nusselt number of the droplet increase with the Reynolds number. On the other hand, the Sherwood number and the Nusselt number exhibit non-monotonic behavior with increasing shock Mach number. Initially, the Sherwood number and the Nusselt number decrease as the shock Mach number is increased from 1.1 till 1.43. As the shock Mach number is increased further, the Sherwood number and the Nusselt number increase. The non-monotonic behavior is due to the transition of the post-shock flow from sub-sonic to the super-sonic speeds as the shock Mach number is increased from 1.1 to 3.5. The surrogate model for Sherwood number developed in this work accounts for the effects of shock-compression of the gaseous phase around the droplet by incorporating shock Mach number within the functional form of the Sherwood number. In contrast with available models in the literature that are commonly used in process scale computations of droplet vaporization (Kreith et al., 2012; Ranz and Marshal, 1952), the current models for the Sherwood number and the Nusselt number are cast as a function of both the shock Mach number and the Reynolds number and will be useful in the macro-scale multi-phase simulations of Aluminumized propellants.

The current models of the Sherwood and the Nusselt numbers are developed with several underlying assumptions. The cylindrical shape of the droplets in 2D is considered to keep the computational cost of the numerical calculations tractable. However, a multifidelity surrogate modelling technique presented in (Das et al., 2018a) can be used to develop a surrogate model for the Sherwood number and the Nusselt number of spherical droplets by utilizing a handful of fully resolved 3D calculation to improve upon the surrogate models obtained from 2D simulations. Furthermore, the chemical reaction between the vaporized aluminum and the air is neglected in the current calculations. The reaction between the aluminum vapor and the air will influence the vaporization rate of the droplet during shock interaction. This effect of chemical reaction on the vaporization rate of the droplets is being investigated in an ongoing work. Simulations of fields of droplets are more desirable than single-droplet simulations but are more expensive; such simulations are also being pursued in ongoing work and will be reported in a future publication.

## Acknowledgments

The authors gratefully acknowledge the financial support from the Air Force Research Laboratory Munitions Directorate (AFRL/RWML), Eglin AFB, under contract number FA8651-16-1-0005 (Program Manager: Dr. David Barrett Hardin).